\documentclass[epj]{webofc}
\usepackage[varg]{txfonts}   % Web of Conferences font
\usepackage{bbold}
\usepackage{hyperref}
\usepackage{xcolor}

\newcommand{\p}{\partial}
\newcommand{\one}{\mathbb{1}}

\newcommand{\Tr}{\mathrm{Tr}}

\newcommand{\UU}{\mathcal{U}}
\newcommand{\WW}{\mathcal{W}}
\newcommand{\CC}{\mathbb{C}}

\newcommand{\der}[2]{\frac{\mathrm{d} #1 }{\mathrm{d} #2 }} % derivative
\newcommand{\diff}{\mathrm{d}} % differential in integrals

\begin{document}
\title{Preserving gauge invariance in neural networks}

\author{\firstname{Matteo} \lastname{Favoni}\inst{1}\fnsep\thanks{\email{favoni@hep.itp.tuwien.ac.at}}
        \firstname{Andreas} \lastname{Ipp}\inst{1}\fnsep\thanks{\email{ipp@hep.itp.tuwien.ac.at}} 
        \firstname{David I.~} \lastname{Müller}\inst{1,2}\fnsep\thanks{\email{dmueller@hep.itp.tuwien.ac.at}} \and
        \firstname{Daniel} \lastname{Schuh}\inst{1}\fnsep\thanks{\email{schuh@hep.itp.tuwien.ac.at}}
        % etc.
}

\institute{Institute for Theoretical Physics, TU Wien, \\
	Wiedner Hauptstr. 8-10, 1040 Vienna, Austria \and
Speaker and corresponding author}

\abstract{
In these proceedings we present lattice gauge equivariant convolutional neural networks (L-CNNs) which are able to process data from lattice gauge theory simulations while exactly preserving gauge symmetry. We review aspects of the architecture and show how L-CNNs can represent a large class of gauge invariant and equivariant functions on the lattice. We compare the performance of L-CNNs and non-equivariant networks using a non-linear regression problem and demonstrate how gauge invariance is broken for non-equivariant models.
}
\maketitle
\section{Introduction} \label{intro}

The success of convolutional neural networks (CNNs) in image recognition has demonstrated that adapting machine learning methods to the specifics of the problem at hand can lead to better performing models with fewer parameters. For example, the problem of image classification (e.g.~determining in a series of images if a particular image shows a certain animal or determining a number shown in an image of hand-written digits) often does not require knowledge about where in the image a certain feature is detected, but just that the image contains it somewhere. In that sense, these problems are invariant under spatial translations. The main idea behind CNNs is to make use of this symmetry by restricting neural network layers to be translationally equivariant. In this context, translational equivariance refers to the fact that applying a spatial translation to an input image yields an appropriately shifted output of the CNN. The concept of equivariance in neural networks has been extended to more general global symmetries (e.g.~rotations, reflections) in the framework of group equivariant CNNs \cite{Cohen:2016aaa} or $G$-CNNs, where $G$ refers to a symmetry group. 
Neural networks have also been generalized to local symmetry, specifically in the case of data defined on curved manifolds~\cite{Cohen:2019aaa}. More generally, geometric deep learning \cite{Bronstein:2021aaa, Gerken:2021sla} is used as an umbrella term for the concept of incorporating the geometry of a machine learning problem in the choice of network architecture. 

Neural network architectures that exhibit symmetry properties are well suited for applications in physics (see e.g.~\cite{Mehta:2019aaa} for a pedagogical introduction), in particular methods that are tailored to preserve the symmetries of physical theories. For example, CNNs (or more generally $G$-CNNs) can be applied to problems in lattice field theory \cite{Zhou:2018ill, Boyda:2020nfh, Blucher:2020mjt, Bachtis:2020ajb, Bulusu:2021rqz, Bachtis:2021xoh, deHaan:2021erb}, which typically exhibit translation, rotation and reflection symmetry. Recently, machine learning models with built-in local symmetry have also been applied in the context of Abelian and non-Abelian lattice gauge theories, e.g.~as generative models for Monte Carlo simulations \cite{Kanwar:2020xzo, Boyda:2020hsi, Albergo:2021vyo}. In these proceedings we review some results of our recent work \cite{Favoni:2020reg} on lattice gauge equivariant convolutional neural networks (L-CNNs), which is a general framework for networks that by construction preserve lattice gauge symmetry in pure SU($N_c$) gauge theory. We first review some notation for lattice gauge theory in Sec.~\ref{lattice} and then introduce the L-CNN in Sec.~\ref{lcnn}. Finally, we demonstrate the performance of L-CNNs compared to non-equivariant CNNs in Sec.~\ref{results}.

\section{Lattice gauge theory} \label{lattice}

The link formalism of lattice gauge theory due to Wilson \cite{Wilson:1974sk} allows us to construct lattice discretizations of non-Abelian Yang-Mills theory with exact lattice gauge symmetry.
In the lattice formulation, gauge fields $A_\mu$ are replaced by (gauge) link variables $U_{x,\mu} \in \mathrm{SU}(N_c)$, defined on the edges of a hypercubic lattice $\Lambda$ with lattice spacing $a$, which connect the starting lattice site $x$ to the end at $x+\mu$.\footnote{We use $x+\mu$ to denote the point $x + a \hat{e}_\mu$, where $\hat{e}_\mu$ is a Euclidean basis vector on the lattice.} For explicitness, we assume the lattice to be finite with periodic boundary conditions. We denote the dimension of the lattice by $D+1$, where $D\geq1$ refers to the spatial dimensions and $\mu = 0$ is the (imaginary) time direction. The size of the lattice is given by $N_t \cdot N_s^D$.
In terms of the gauge field $A_\mu(x) \in \mathfrak{su}(N_c)$, a gauge link $U_{x,\mu}$ is given by the path-ordered exponential\footnote{Here we use the convention that the path ordering operator $\mathcal{P}$ shifts fields earlier in the path to the left, and fields later to the right of the product, i.e.~$\mathcal{P}(O(a)O(b)) = O(a)O(b)$ if $a<b$ and $\mathcal{P}(O(a)O(b)) = O(b)O(a)$ if $b<a$.}
\begin{align}
    U_{x,\mu} = \mathcal{P} \exp{\left( i \intop^1_0 \diff s \der{x^\nu(s)}{s} A_\nu(x(s))  \right)} \, ,
\end{align}
where $x(s): [0,1] \rightarrow \mathbb{R}^4$ defines the straight-line path connecting $x$ to $x+\mu$, and $\mathbb{R}^4$ is the four-dimensional Euclidean spacetime.  The geometric interpretation of gauge links is that they describe parallel transport along the edges of the lattice.
Under general gauge transformations $\Omega: \mathbb{R}^4 \rightarrow \mathrm{SU}(N_c)$ of the gauge field,
\begin{align}
    A_\mu(x) \rightarrow \Omega(x) \left( A_\mu(x) - i \partial_\mu \right) \Omega^\dagger(x),
\end{align}
the gauge links transform according to
\begin{align} \label{eq:U_trans}
    U_{x,\mu} \rightarrow \Omega_x U_{x,\mu} \Omega^\dagger_{x+\mu}.
\end{align}
For concreteness, we use the fundamental representation of $\mathfrak{su}(N_c)$ and  SU($N_c$) to represent $A_\mu$ and $U_{x,\mu}$ as complex matrices. The inverse link is denoted by $U_{x+\mu, -\mu} = U^\dagger_{x,\mu}$. In the limit of small lattice spacing $a \rightarrow 0$, gauge links can be approximated with the matrix exponential $U_{x,\mu} \simeq \exp{(i a A_\mu(x+\frac{1}{2} \mu))}$ at the mid-point $x+\frac{1}{2} \mu$. 

Multiple links connecting consecutive points can be multiplied to form Wilson lines along arbitrary paths on the lattice, and, in particular, closed paths or Wilson loops, where the start and end point coincide. The simplest loop is the $1 \times 1$ loop called plaquette
\begin{align}
    U_ {x,\mu\nu} = U_{x,\mu} U_{x+\mu, \nu} U_{x+\mu+\nu, -\mu} U_{x+\mu, -\mu},
\end{align}
which under gauge transformations transforms according to
\begin{align}
    U_{x,\mu\nu} \rightarrow \Omega_x U_{x,\mu\nu} \Omega^\dagger_x.
\end{align}
In the continuum limit, the plaquette approximates the non-Abelian field strength tensor $F_{\mu\nu} = \p_\mu A_\nu - \p_\nu A_\mu - i \left[A_\mu, A_\nu\right]$ via $U_{x,\mu\nu} \simeq \exp{\left( i a^2 F_{\mu\nu} \right)}$. The Yang-Mills action can then be approximated by the Wilson action \cite{Wilson:1974sk}
\begin{align} \label{eq:wilson_action}
S_W[U] = \frac{2}{g^2} \sum_{x \in \Lambda} \sum_{\mu < \nu} \mathrm{Re} \, \Tr \left[ \one - U_{x,\mu\nu} \right],
\end{align}
where $g > 0$ is the Yang-Mills coupling constant. Because of the trace over locally transforming plaquettes, the Wilson action is invariant under general lattice gauge transformations. This gauge invariant action can be used to perform Monte Carlo sampling of the path integral and compute expectation values of observables in lattice QCD at finite temperature \cite{Wilson:1980a}.

\section{Lattice gauge equivariant convolutional neural networks} \label{lcnn}

This section reviews some aspects of the L-CNN architecture \cite{Favoni:2020reg}. Similar to conventional CNNs, we construct our architecture from elementary layers, which explicitly respect lattice gauge symmetry. First, we define the data points for L-CNNs as tuples $(\UU, \WW)$, where $\UU = \{ U_{x,\mu} \}$ is the set of gauge links of a particular lattice configuration, and $\WW = \{ W_{x,i} \}$ with $W_{x,i} \in \CC^{N_c \times N_c}, \, \forall x \in \Lambda, \, 1 \leq i \leq N_\mathrm{ch}$ is a set of locally transforming complex matrices with $N_\mathrm{ch} \in \mathbb{N}$ channels. Under lattice gauge transformations, we require that the matrices $\WW$ transform locally
\begin{align} \label{eq:W_trans}
    W_{x,i} \rightarrow \Omega_x W_{x,i} \Omega^\dagger_x.
\end{align}
Comparing L-CNNs to CNNs, the $\WW$ matrices can be thought of as feature maps, while the links $\UU$ provide the geometrical information to compare data at different lattice sites via parallel transport. We consider data points that are related by gauge transformations to be gauge equivalent.

Next, we need to specify what data the $\WW$ matrices represent. In practice (see Sec.~\ref{regression}) we use plaquettes $U_{x,\mu\nu}$ at each point $x$ on the lattice as input $\WW$'s such that each face defined by the pair $(\mu , \nu)$ is assigned to a particular channel. More generally, it is also possible to include all Polyakov loops (which are non-contractible loops wrapping around the periodic boundary of the lattice) at each lattice site in the input. We show why this is a particularly useful choice for input data in Sec.~\ref{proof}, where we prove that arbitrary contractible and non-contractible Wilson loops can be generated with L-CNNs using just two elementary layers. 

\subsection{Equivariant convolutions and bilinear layers}

We introduce two gauge equivariant operations acting on data points $(\UU, \WW)$: lattice gauge equivariant convolutions (L-Convs) and lattice gauge equivariant bilinear layers (L-Bilin). We formulate the 
layers in such a way that the output of these operations transforms consistently under Eqs.~\eqref{eq:U_trans} and \eqref{eq:W_trans}.

The L-Conv operation maps a data point $(\UU, \WW)$ to a new data point $(\bar \UU, \bar \WW)$  given by
\begin{align} \label{eq:lconv}
\bar U_{x,\mu} &= U_{x,\mu }, \nonumber \\
\bar W_{x, i} &= \sum_{j, \mu, k} \omega_{i, j, \mu, k} U_{x, k\cdot \mu} W_{x + k\cdot \mu, j} U^\dagger_{x, k\cdot \mu},
\end{align}
where $\omega_{i, j, \mu, k} \in \CC$ are the trainable weight parameters (or kernel weights) with indices for output channels $1 \leq i \leq N_\mathrm{ch, out}$, input channels $1 \leq j \leq N_\mathrm{ch, in}$, lattice directions $0 \leq \mu \leq D$, and distances $-K \leq k \leq K$. Here, $K \in \mathbb N$ is the size of the kernel, which determines the receptive field, i.e.~how many lattice points are considered when computing the convolution. The $\WW$ matrices at different lattice sites are parallel transported to the common point $x$ along straight paths using the Wilson lines constructed from links:
\begin{align}
    U_{x,k\cdot\mu} = \prod^{k-1}_{i=0} U_{x+i \cdot \mu, \mu} = U_{x,\mu} U_{x+\mu, \mu} U_{x+2\cdot\mu,\mu} \dots U_{x+(k-1)\cdot \mu, \mu}.
\end{align}
Under gauge transformations, Eqs.~\eqref{eq:U_trans} and \eqref{eq:W_trans}, we find
\begin{align}
\bar W_{x,i} \rightarrow \sum_{j, \mu, k} \omega_{i, j, \mu, k} \Omega_x U_{x, k\cdot \mu} \Omega^\dagger_{x+k\cdot \mu} \Omega_{x+k,\mu} W_{x + k\cdot \mu, j} \Omega^\dagger_{x+k\cdot \mu} \Omega_{x+k\cdot \mu} U^\dagger_{x, k\cdot \mu} \Omega^\dagger_x = \Omega_x \bar W_{x,i} \Omega^\dagger_x,
\end{align}
which shows that gauge equivariance is satisfied. Analogous to convolutional layers in standard CNNs, the output of L-Conv is also equivariant under lattice translations.

The second layer we introduce is a local bilinear operation. The L-Bilin layer maps two input data points $(\UU, \WW)$ and $(\UU', \WW')$ with $\UU' = \UU$ to a new data point $(\bar \UU, \bar \WW)$
\begin{align} \label{eq:lbilin}
\bar U_{x,\mu} &= U_{x,\mu }, \nonumber \\
\bar W_{x,i} &= \sum_{j,k} \alpha_{i,j,k} W_{x,j} W'_{x,k},
\end{align}
where $\alpha_{i,j,k} \in \CC$ are trainable weights with indices $1 \leq i \leq N_\mathrm{ch, out}$, $1 \leq j \leq N_\mathrm{ch, in}$, and $1 \leq k \leq N'_\mathrm{ch, in}$. The bilinear operation multiplies two locally transforming matrices at the same lattice site, which guarantees gauge equivariance:
\begin{align}
\bar W_{x,i} \rightarrow \sum_{j,k} \alpha_{i,j,k} \Omega_x W_{x,j} \Omega^\dagger_x \Omega_x W'_{x,k} \Omega^\dagger_x = \Omega_x \bar W_{x,i} \Omega^\dagger_x.
\end{align}
As in the case of L-Conv, L-Bilin is equivariant under translations as well. The bilinear operation reduces to a quadratic layer when we use the same data point for both arguments, i.e.~$(\UU, \WW) = (\UU', \WW')$.

Both operations only modify the $\WW$ part of the tuple $(\UU, \WW)$. It is also possible to formulate equivariant layers that modify the links $\UU$ (see e.g.~L-Exp in Ref.~\cite{Favoni:2020reg}), but for the purposes of these proceedings we only focus on L-Conv and L-Bilin. We also note that the expressivity of these layers can be further increased by adding additional channels to $\WW$ prior to the operation. Specifically, before applying L-Conv or L-Bilin one may add unit elements and hermitian conjugates:
\begin{align}
     (W_{x,1}, W_{x,2}, \dots, W_{x,N_\mathrm{ch,in}}) \rightarrow (\one, W_{x,1}, W_{x,2}, \dots, W_{x,N_\mathrm{ch,in}}, W^\dagger_{x,1}, W^\dagger_{x,2}, \dots, W^\dagger_{x,N_\mathrm{ch,in}}).
\end{align}
This increases the number of channels from $N_\mathrm{ch}$ to $2N_\mathrm{ch}+1$. The unit elements generate bias terms for L-Conv, and bias and linear terms for L-Bilin. Hermitian conjugates correspond to changing the orientation of input plaquettes or Polyakov loops.

\subsection{Generating arbitrary loops} \label{proof}

\begin{figure}
    \centering
    \includegraphics[scale=1.1]{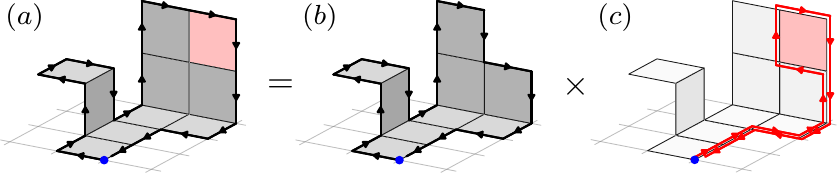}
    \caption{An arbitrary contractible Wilson loop on the lattice consisting of $n$ squares (a) can be decomposed into a product of a loop with $n-1$ squares missing one plaquette (b) and a parallel transported single plaquette (c). Figure from \cite{Favoni:2020reg}.}
    \label{fig:proof}
\end{figure}

It is important to address which class of functions can be represented by neural network architectures. Universality theorems exist for fully connected neural networks \cite{Lu:2017a} and deep CNNs \cite{Zhou:2018a}, which demonstrate that neural networks are universal function approximators. In order to establish that L-CNNs are able to represent a large class of gauge equivariant functions, we prove by induction that arbitrary contractible and non-contractible Wilson loops can be constructed from repeated applications of L-Conv and L-Bilin. 

Consider a data point tuple $(\UU, \WW)$, where the set of $\WW$ matrices are populated with all possible plaquettes at each lattice site. Initializing $\WW$ in this manner trivially generates all possible $1 \times 1$ loops or loops of size $1$. Next, consider an arbitrary contractible loop of size $n$ (i.e.~it consists of $n$ squares) on the lattice, see Fig.~\ref{fig:proof} (a). Such a loop can be factorized into a loop of size $n-1$ (Fig.~\ref{fig:proof} (b)) with one missing plaquette and a transported single plaquette where the path traces along parts of the boundary of the original loop (Fig.~\ref{fig:proof} (c)). The factorization here refers to the composition of paths, which in the case of Wilson loops is realized by multiplying their matrix representations. L-CNNs can realize this factorization explicitly: Stacks of multiple L-Conv layers can generate arbitrary parallel transported plaquettes. For example, the parallel transported plaquette in Fig.~\ref{fig:proof} (c) can be generated with four L-Conv layers. On the other hand, the multiplication of the loop of size $n-1$ and the single transported plaquette can be realized by a single L-Bilin layer. By induction, every contractible loop can be constructed with L-Conv and L-Bilin operations starting from elementary plaquettes. This construction can be generalized to non-contractible loops (which wrap around the boundary of the lattice) by including, in addition to plaquettes, all possible straight line Polyakov loops  at every lattice site in the initial set of $\WW$ matrices.

Generally, L-CNNs consisting of stacks of L-Conv and L-Bilin layers can represent arbitrary linear combinations of Wilson loops of various shapes and sizes. The largest possible size and shape is determined by the kernel sizes and the number of layers.
By including gauge equivariant non-linear activation functions in the L-CNN architecture (as introduced in our paper \cite{Favoni:2020reg}), it is possible to represent non-linear functions of loops as well. Finally, if the desired output of an L-CNN is supposed to be gauge invariant (e.g.~in the case of a regression problem for gauge invariant observables), it is possible to simply compute the trace of all matrices in $\WW$.
This renders the output of an L-CNN invariant because of the cyclic property of the trace. 

\section{Results and discussion} \label{results}

In order to test our new architecture, we designed a non-linear regression problem and tried out various L-CNN and non-equivariant CNN (baseline) architectures to solve it. The performance of 
L-CNNs and baseline networks are compared, and we investigate the breaking of gauge symmetry in baseline CNNs using adversarial attacks.

\subsection{Wilson loop regression} \label{regression}

\begin{figure}
    \centering
    \includegraphics[width=\textwidth]{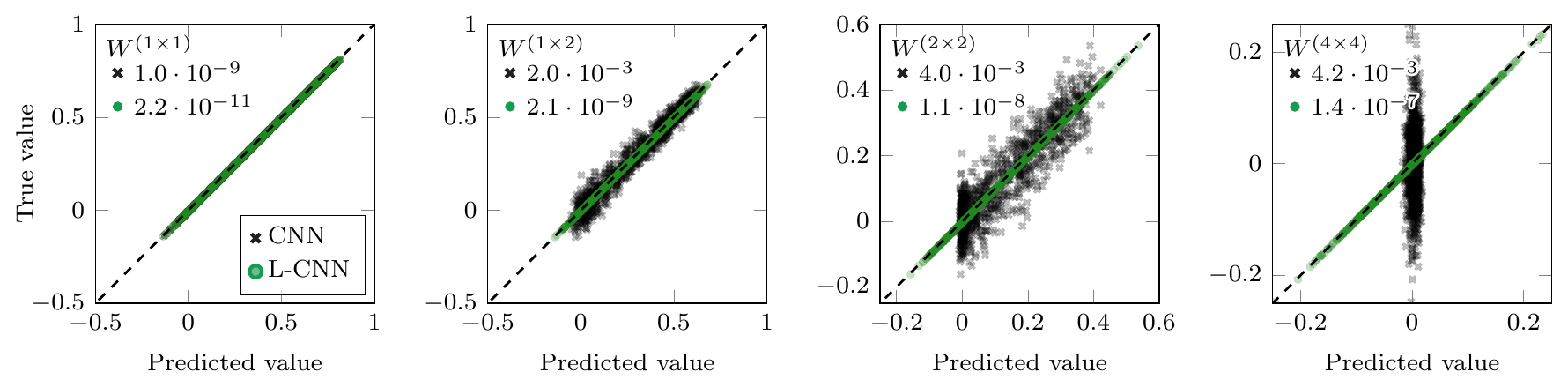}
    \caption{Scatter plots of our best L-CNN and baseline CNN models according to validation MSE on $8\cdot 8$ lattices. True values of the Wilson loop are plotted against the predictions of the L-CNN model (green dots) and the baseline CNN (black crosses). High accuracy is achieved when all points are distributed close to the $45^\circ$ line. The MSE is stated in the upper left corner of each panel. We observe that our best L-CNN models achieve low MSE for all studied loop sizes, while baseline CNNs perform worse at larger loop sizes. Figure adapted from \cite{Favoni:2020reg}. }
    \label{fig:scatter}
\end{figure}

The proposed regression problem consists of two-dimensional ($D=1$) lattice configurations sampled from an SU(2) Monte Carlo simulation, from which networks should compute the real value of the trace of $n \times m$ Wilson loops, i.e.~
\begin{align}
W^{(m \times n)}_{x,\mu\nu} = \frac{1}{N_c} \mathrm{Re}\,  \Tr \left[ U^{(m \times n)}_{x,\mu\nu} \right],
\end{align}
which is a gauge invariant observable. In our computational experiments we use $1 \times 1$ (as a trivial example), $1 \times 2$, $2 \times 2$ and $4 \times 4$ loops. We sample configurations for various values of the coupling constant on an $N_t \cdot N_s = 8 \cdot 8$ lattice to create the training and validation set. Test sets are generated for lattice sizes up to $64 \cdot 64$. For each architecture, trainable weights are randomly initialized multiple times and independently trained. We use mean squared error (MSE) as the objective (or loss) function to optimize during training. Both \mbox{L-CNNs} and baseline CNNs are provided with links and plaquettes in the input layer. For more details regarding architectures and the training process, we refer to our original paper and its forthcoming supplementary materials \cite{Favoni:2020reg}.

Summarizing the main results of our baseline study, we find that L-CNNs are indeed able to solve the regression problem for all sizes of the loop to a high degree of accuracy and also exhibit generalization to larger lattices. Baseline CNNs typically are not able to find adequate solutions, and the quality of their fits deteriorates with increased loop size. In Fig.~\ref{fig:scatter} we show scatter plots for the best models of each type, where true values (i.e.~labels in the dataset) are plotted against the predictions made by the networks. 

\subsection{Adversarial attacks for non-equivariant networks}

\begin{figure}
    \centering
    \includegraphics[scale=0.9]{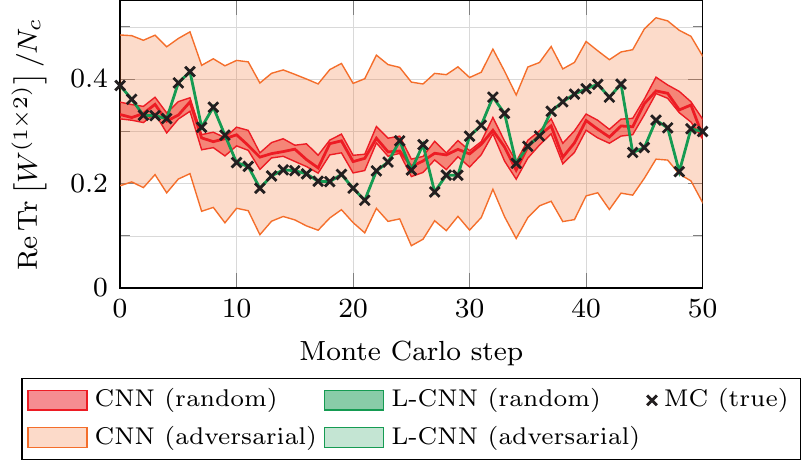}
    \caption{Uncertainty in predictions due to broken gauge symmetry in L-CNNs (green bands) and baseline CNNs (red bands) on $8\cdot8$ test data for the $1 \times 2$ loop regression problem. The true values (labels) are shown as black crosses. We use the same models as in Fig.~\ref{fig:scatter}. L-CNNs are invariant by construction and therefore only show deviations on the order of numerical precision. Baseline CNNs show much larger deviations, where the dark red band corresponds to errors due to random transformations, and the light red band shows the result of adversarial attacks. Figure from \cite{Favoni:2020reg}.}
    \label{fig:uncertainty}
\end{figure}

While our L-CNN architectures are gauge invariant by construction (up to numerical precision), the baseline CNNs must learn this symmetry during the training process. Therefore, it is expected that even well-trained baseline CNNs at best only exhibit approximate gauge symmetry. The extent to which gauge symmetry is broken can be studied by applying gauge transformations to the input layer
\begin{align}
    U_{x,\mu} \rightarrow \Omega_x U_{x,\mu} \Omega_{x+\mu}^\dagger, \qquad W_{x,i} \rightarrow \Omega_x W_{x,i} \Omega_x^\dagger,
\end{align}
with $\Omega_x \in \mathrm{SU}(N_c)$ and determining how much the predictions of a non-equivariant model change as a result. For a given model and a given lattice configuration, we apply two types of transformations: multiple random gauge transformations, where each $\Omega_x$ is sampled randomly, and adversarial attacks, where $\Omega_x$ is optimized to produce the largest deviation between the original prediction and the gauge transformed prediction. The adversarial attack is 
unique to each configuration and model, while random transformations do not depend on either. We show the results of such an attack for our best L-CNN and baseline CNN models for the $1 \times 2$ loop in Fig.~\ref{fig:uncertainty}. The L-CNN model is entirely unaffected by transformations (up to numerical precision), but the predictions of the baseline network show large errors in the case of adversarial attacks.

\section{Conclusions}

In these proceedings we have reviewed some aspects of the L-CNN architecture, particularly the L-Conv and L-Bilin layer, which in combination can be used to generate arbitrarily shaped Wilson loops on the lattice. We have also presented comparisons between L-CNNs and non-equivariant CNNs using a non-linear regression task for Wilson loops and demonstrated the breaking of gauge invariance in non-equivariant networks. 

In this work we have mainly focused on the formulation and some properties of L-CNNs, but from a practical perspective, it would be interesting to apply the L-CNN architecture to normalizing \cite{Kanwar:2020xzo, Boyda:2020hsi, Albergo:2021vyo} and continuous \cite{deHaan:2021erb} flow models. On the other hand, in order to provide a more solid mathematical foundation, it would be worthwhile to understand L-CNNs as a special case of CNNs on principal bundles in the vein of gauge equivariant CNNs for curved manifolds \cite{Cohen:2019aaa} or gauge equivariant convolutions defined in \cite{Gerken:2021sla}. 

\vspace{1em}

\begin{acknowledgement}
DM thanks Jimmy Aronsson for many helpful discussions regarding equivariance, neural networks and geometric deep learning. This work has been supported by the Austrian Science Fund FWF No.~P32446-N27, No.~P28352 and Doctoral program No. W1252-N27. The Titan\,V GPU used for this research was donated by the NVIDIA Corporation.
\end{acknowledgement}

\bibliography{refs}

\end{document}